\documentstyle[preprint,revtex]{aps}
\begin{document}
%
\font\fortssbx=cmssbx10 scaled \magstep2
\hbox{ 
$\vcenter{\fortssbx University of Wisconsin - Madison}$ }
\hfill\vbox{\hbox{\bf MAD/PH/704}\hbox{June 1992}}\par
\vspace{.5in}
\begin{title}
What $e^+e^-$ collider could make a ``no-lose'' search\\ for MSSM Higgs bosons?
\end{title}
\author{\mbox{\llap{V.~Barger\rlap,$^*$ Kingman~Cheung\rlap,$^*$}
\rlap{R.~J.~N.~Phillips$^\dagger$ and A.~L.~Stange$^*$}\phantom{999}}}
\begin{instit}
$^*$Physics Department, University of Wisconsin, Madison, WI 53706, USA\\
$^\dagger$Rutherford Appleton Laboratory, Chilton, Didcot, Oxon, OX11 0QX
England
\end{instit}

\thispagestyle{empty}

\begin{abstract}
\nonum\section{abstract}
The lightest CP-even Higgs boson $h$ in the minimum supersymmetric
standard model (MSSM) has a mass upper bound depending on the top quark
and squark masses. An $e^+e^-$ collider with enough energy and luminosity
to produce $h+Z$ at measurable rates up to the maximum $h$ mass would cover
the entire MSSM parameter space, if $h+A$ production was also
searched for. We explore the energy and/or luminosity  needed for
various top quark and squark masses. For $m_t=150$\,GeV and 1\,TeV SUSY mass
scale, a 230\,GeV collider with 10\,fb$^{-1}$ luminosity would suffice.
\end{abstract}

\newpage

The theoretical appeal of supersymmetry (SUSY) is that it solves the problem of
large radiative corrections in the scalar sector, associated with the grand
unification scale. The minimal supersymmetric extension of the Standard Model
(MSSM)~\cite{hhguide} has five Higgs bosons, one of which ($h$) is necessarily
relatively light; their discovery could contribute the first direct evidence
both for SUSY and for the Higgs mechanism. These Higgs bosons are therefore the
object of intense experimental investigation; a lower limit $m_h\agt40$\,GeV
has already been set by $e^+e^-$ experiments at
LEP\,I~\cite{aleph,delphi,L3,opal} and the range of search will be extended at
LEP\,II  with CM energy $\sqrt s=190\,$GeV to 240\,GeV possible~\cite{rubbia}.
In this letter we address the question: what is the lowest energy $e^+e^-$
collider that could completely cover the MSSM parameter
space~\cite{kz,680,baer,gunion,696} and thereby independently guarantee
discovery or rejection of the MSSM? This question is relevant because LEP\,I,
LEP\,II, SSC and LHC will not fully cover all MSSM
parameters~\cite{kz,gunion,696}, and the possibilities of higher energy
$e^+e^-$ linear colliders are being examined~\cite{janot}.

The Higgs sector of the minimum supersymmetric standard model (MSSM) has three
neutral and two charged Higgs bosons, $h,\ H,\ A,\ H^\pm$ of which $h$ and $H$
are CP-even and $m_h<m_H$; a mixing angle $\alpha$ appears in the $h$ and $H$
couplings. At tree level all their masses and couplings are controlled by two
parameters, that may be taken to be $m_A$ and the ratio $\tan\beta=v_2/v_1$ of
vacuum expectation values giving masses to up-type quarks ($v_2$) and down-type
quarks ($v_1$) respectively; renormalization group arguments in no-scale
models~\cite{zwirner}
suggest that $1<\tan\beta<m_t/m_b$ but $m_A$ is unconstrained. At one-loop
level, however, there are significant radiative corrections~\cite{loopcor},
that depend on several other parameters but especially on the top quark and
squark masses; as a result the $h$ mass has an upper bound
\begin{equation}
m_h^2 \alt M_Z^2\cos^2\beta
+ {6G_F\over \pi^2 \sqrt 2} m_t^4
\ln\left(\tilde m\over m_t\right) \;, \label{bound}
\end{equation}
in an approximation where the usual SUSY parameters $A_t,\ A_b$ and $\mu$ are
set to zero and $\tilde m$ is the common SUSY mass scale. The masses of $H$,
$A$ and $H^\pm$ have no upper bounds. In our present discussion we shall use
non-zero values of all SUSY parameters, following Ref.~\cite{680}, with
$\tilde m\simeq 1$\,TeV and $1<\tan\beta<30$. The important parameter is
the shift in the $m_h$ upper bound; we note that large changes in $\tilde m$
are effectively equivalent here to small changes in $m_t$.

At $e^+e^-$ colliders the signals for Higgs bosons are relatively clean and the
opportunities for discovery and detailed study will be excellent. The principal
production channels are
\begin{mathletters}
\begin{eqnarray}
e^+e^- &\to& Z \to Zh, ZH, Ah, AH \;, \label{channels a} \\
e^+e^- &\to& \nu\bar\nu W^*W^* \to \nu\bar\nu h, \nu\bar\nu H \;,
\label{channels b}\\
e^+e^- &\to& t\bar th, t\bar tH, t\bar tA \;,\label{channels c}\\
e^+e^- &\to& Z,\gamma \to H^+H^- \;, \label{channels d}\\
e^+e^- &\to& t\bar t \to b\bar b \, H^+(W^+) \, H^-(W^-) \;.\label{channels e}
\end{eqnarray}
\end{mathletters}
The $s$-channel processes (\ref{channels a}) offer the biggest contributions at
the lower energies.
In the limit $m_A\to\infty$ we have $m_H\simeq m_{H^+}\simeq m_A$ while $m_h$
approaches the upper bound of Eq.~(\ref{bound}) and only $h$ can be produced at
any given collider. Thus the channel $e^+e^-\to Zh$ and its kinematical limits
are critical in any complete search of MSSM parameter space. The $Zh$
production cross section contains an overall factor $\sin^2(\beta-\alpha)$
which suppresses it in certain parameter regions (with $m_A<100\,$GeV and
$\tan\beta$ large); fortunately the $Ah$ production cross section contains the
complementary factor $\cos^2(\beta-\alpha)$. Hence the $Zh$ and $Ah$ channels
together are well suited to cover all regions in the $(m_A,\,\tan\beta)$ plane,
provided that the CM energy is high enough for $Zh$ to be produced through the
whole $m_h$ mass range, and that an adequate event rate can be achieved. These
conditions are already shown to be fulfilled~\cite{janot} for $\sqrt s
=500$\,GeV with assumed luminosity 10\,fb$^{-1}$. In the present work we study
how well these conditions can be fulfilled at lower energies with various
luminosities.

Our discussion centers on the $s$-channel production channels $e^+e^-\to
Zh,Ah,ZH,AH$, neglecting all others for simplicity (although other channels
would obviously contribute to an eventual search and analysis). We also
consider only the decays $Zh,Ah,ZH,AH\to\tau\tau jj$ (where $j$ denotes a
$b$-jet); this generally has substantial branching fractions at least in the
$Zh$ and $Ah$ cases. We rely here on the possibility of recognizing and
kinematically reconstructing $\tau$ jets
experimentally~\cite{aleph,delphi,L3,opal,janot}; no $b$-tagging of the other
jets is assumed. This approach is conservative, since it implicitly ignores
substantial $Z\to\ell^+\ell^-, \nu\bar\nu$ decay modes that could enhance the
detectability of $Zh$ and $ZH$ production. Since
the $\ell^+\ell^- jj$ and $\nu\bar\nu jj$ channels have smaller
signal/background ratios than the $\tau\tau jj$ channel,  the net
significance of $Zh$ or $ZH$ signals would not be dramatically increased by
including the former channels.

For any given energy and MSSM parameters, we calculate the $Zh$, $Ah$, $ZH$ and
$AH$ production cross sections and decay branching fractions from standard
formulas~\cite{hhguide} with one-loop corrections as described in
Ref.~\cite{680}. We omit bremsstrahlung and beamstrahlung corrections,
that are not very large in currently favored collider designs.
It is assumed that the Higgs bosons do not decay to light SUSY
particles. The signals take the form of peaks in the distributions of
invariant mass $m(\tau\tau)$ and $m(jj)$, centered at values $m_h$, $m_H$ and
$m_A$, with an associated peak at $M_Z$ also. These two $m(\tau\tau)$ and
$m(jj)$ distributions are added to enhance the statistics, thus giving two
counts per event. An irreducible background from $e^+e^-\to ZZ$ production and
decay has a peak centered at $M_Z$; all other backgrounds can however be
suppressed by suitable cuts at little cost to the signals~\cite{janot}. For
present purposes we estimate the background from the numerical simulation of
Ref.~\cite{janot}, scaling the number of events according to the assumed
luminosity and the energy-dependence of the $e^+e^-\to ZZ\to\tau\tau jj$ cross
section with $|\cos\theta|<0.9$ for the $\tau$'s and jets. We assume that the
acceptances of the $Zh\,(ZH)$ and $Ah\,(AH)$ signals  remain 46\% and 52\%,
respectively, as in Ref.~\cite{janot} and that the Higgs boson peaks have the
same mass resolution as the $Z$ peak. This approach is approximate, but
avoids lengthy Monte Carlo simulations for each of the many different energies
and parameter settings that we have to consider.

For each input set of SUSY parameters, CM energy $\sqrt s$ and integrated
luminosity $\cal L$, we define the signals, backgrounds and discovery criteria
of the MSSM mass-peaks as follows.
 For an isolated peak, the signal strength $S$ is taken to be the
expected number of signal counts falling in a 10~GeV mass bin centered at the
corresponding Higgs boson mass. When two Higgs peaks approach within 10~GeV we
combine them; the signal strength $S$ is then the
total number of counts expected in a 10~GeV bin centered at the weighted mean
mass. The background strength $B$ is taken to be the total number of $Z$-decay
counts (both from $ZZ$ and from $Zh,ZH$ production with the resolution of
Ref.~\cite{janot}) falling in the same mass bin. If the signal bin center is
separated by more than 5\,GeV from $M_Z$, our discovery criteria are
$S\Big/\sqrt B>4$ with $S>4$ counts. With such a separation, we expect that a
distinct peak will be seen or that a recognizable distortion of the $Z$ peak
will be evident. But if the separation from $M_Z$ is less than 5~GeV, we can
only infer
the presence of a new signal if the height of the supposed $Z$ peak differs
substantially from the expected $ZZ$ background contribution. In this latter
case we rely entirely on normalization and therefore require a higher degree of
significance. Here the signal $S$ is defined to be the sum of the MSSM
 ($h$, $H$, $A$ and $Z$) contributions falling in a 10~GeV bin centered at
$M_Z$, and $B$ is the expected $ZZ$ background in the same bin; in this case we
define a discoverable signal to have $S\Big/\sqrt B>6$ with $S>5$ counts.

In principle, $b$-tagging of the quark jets offers another way to distinguish
the presence of a Higgs boson contribution hiding under the $Z$ peak in the
$\tau\tau jj$ channel. With typical LEP-type
microvertex tagging, it should be possible to achieve say 38\% tagging
efficiency for Higgs${}\to b\bar b$ pairs coupled with 11\% tagging efficiency
for $Z\to jj$ pairs. If we consider just the $m(jj)$ distribution for
simplicity, a tagged signal would then have significance increased by
approximately the factor $0.38\big/\sqrt{0.11\,}=1.14$ above that of the
corresponding untagged signal (discussion of the $m(\tau\tau)$ distribution is
more complicated). Greater enhancement may well be attainable in the future. It
may also be possible to establish a signal in the ratio of tagged/untagged
events, but this seems to require much higher statistics to achieve serious
significance. Thus $b$-tagging appears to offer real but not dramatic
improvements in sensitivity in the future; conservatively we neglect it here.

For full coverage of
the $(m_A,\,\tan\beta)$ plane, the CM energy should be about
10\,GeV or more above the maximum $Zh$ threshold,
\begin{equation}
\sqrt s\,({\rm threshold}) = m_h\,({\rm max}) + M_Z \;,
\end{equation}
where $m_h\,({\rm max})$ is the largest value of $m_h$ in Eq.~(\ref{bound}).
For $\tilde m=1$\,TeV this threshold is 207\,GeV for $m_t=150$\,GeV and
240\,GeV for $m_t=200$\,GeV (the highest value of $m_t$ allowed by analyses of
radiative corrections\cite{langacker92}). Apart from these threshold
considerations, the principal factors that determine the discovery regions
(where one or more MSSM signals are detectable) in the $(m_A,\,\tan\beta)$
plane are luminosity $\cal L$, top quark mass $m_t$ and CM energy $\sqrt s$.
Figure~\ref{limits} illustrate the effects of these
factors separately, by means of four examples.

\begin{itemize}

\item[(i)] Luminosity: in Figs.~\ref{limits}(a),(b) we hold $m_t=150$\,GeV and
$\sqrt s=215$\,GeV fixed (with all SUSY parameters fixed as in
Ref.~\cite{680}), and compare discovery limits for
${\cal L}=1$~fb$^{-1}$  and  ${\cal L}=10$~fb$^{-1}$,
corresponding respectively to one month and one year running
at $L=10^{33}/\rm cm^2/s$.  Here the  $(m_A,\,\tan\beta)$
plane is fully accessible kinematically, but good luminosity
is still needed to guarantee discovery; in fact
${\cal L} \agt 20$~fb$^{-1}$  would give full coverage.

\item[(ii)] Top quark mass: in Figs.~\ref{limits}(b),(c) we hold
${\cal L}=10$~fb$^{-1}$ and $\sqrt s=215$\,GeV  fixed and
compare the discovery limits for  $m_t=150$\,GeV and
$m_t=200$\,GeV.  Coverage becomes easier as $m_t$
decreases; there would be complete coverage in this case
with $m_t \alt 120$\,GeV.

\item[(iii)] CM energy: in Figs.~\ref{limits}(c),(d) we hold
${\cal L}=10$~fb$^{-1}$
and $m_t=200$\,GeV fixed and compare the discovery limits at
$\sqrt s=215$ and $270$\,GeV.  We see that increasing $s$ in this
range generally widens the accessible region, although this
is not uniformly true since the signals have different energy
dependences in different parts of the plot.  In fact, with our
discovery criteria it appears that complete coverage is not
achieved at any energy with this particular choice of
${\cal L}$ and $m_t$.  We remark in passing that the
small area lower left, inaccessible in Figs.~\ref{limits}(a) and
\ref{limits}(d), is not well served by $\tau\tau jj$ signals, since $h\to AA$
dominates the $h$ decays here; however, this region is already excluded by
LEP\,I data~\cite{aleph,delphi,L3,opal}.

\end{itemize}

The final question is, what combinations of collider parameters $\sqrt s$ and
$\cal L$  would just achieve complete coverage of the $(m_A,\,\tan\beta)$ plane
for given $m_t$? Figure~\ref{s,L plane} shows the limiting curves in the
$(\sqrt s,\,{\cal L})$ plane, for various values of $m_t$; we recall that
changes in $\tilde m$ can be effectively absorbed into $m_t$, and that
LEP\,I searches have already excluded small $m_A$
values~\cite{aleph,delphi,L3,opal}. Pairs of values $(\sqrt s,\,{\cal L})$ that
lie above the limiting curves have ``no-lose'' discovery potential in the MSSM,
according to our approximations.  For example, with $m_t = 150$ GeV and a 1 TeV
SUSY mass scale, a 230 GeV $e^+e^-$ collider with 10 fb$^{-1}$ luminosity would
suffice.

\newpage

We thank Peter Norton for conversations about experimental issues and Paul
Dauncey for information about $b$-tagging efficiencies.
This work is supported in part by the U.S.~Department of Energy under contract
No.~DE-AC02-76ER00881, in part by the Texas National Research Laboratory
Commission under grant No.~RGFY9173, and in part by the University of Wisconsin
Research Committee with funds granted by the Wisconsin Alumni Research
Foundation.


\figure{\label{limits}
Discovery limits in the $(m_A,\,\tan\beta)$ plane for $e^+e^-\to Zh,Ah,ZH,AH$
signals in the $\tau\tau jj$ channel, for various illustrative cases:
(a)~$\sqrt s=215{\rm\,GeV,}\ m_t=150{\,\rm GeV},\ {\cal L}=1\rm\,fb^{-1}$;
(b)~$\sqrt s=215{\rm\,GeV,}\ m_t=150{\,\rm GeV},\ {\cal L}=10\rm\,fb^{-1}$;
(c)~$\sqrt s=215{\rm\,GeV,}\ m_t=200{\,\rm GeV},\ {\cal L}=10\rm\,fb^{-1}$;
(d)~$\sqrt s=270{\rm\,GeV,}\ m_t=200{\,\rm GeV},\ {\cal L}=10\rm\,fb^{-1}$.}

\figure{\label{s,L plane}
Conditions for covering the whole MSSM $(m_A, \tan\beta)$ plane
with $m_A \le 1$\,TeV and $1<\tan\beta<30$. Limiting
curves are shown in the $(\sqrt s,\,{\cal L})$ plane, for various values of
$m_t$. The region that does not give complete coverage for $m_t=150$\,GeV is
shaded.}


\begin{references}

\bibitem{hhguide}
For a recent reviews see J.~Gunion, H.~Haber, G.~Kane
and S.~Dawson, {\it The Higgs Hunter's Guide}, (Addison-Wesley, 1990);
X.~Tata in {\it The Standard Model and Beyond}, edited by J.~E.~Kim
        (World Scientific, 1991) p.~379.

\bibitem{aleph}
ALEPH Collaboration, D.~Decamp {\it et al.}, Phys.\ Lett.\ {\bf B265},
475 (1991), and CERN-PPE/91-149 to be published in Physics Reports.

\bibitem{delphi}
DELPHI Collaboration: P.~Abreu {\it et al.}, Phys.\ Lett.\ {\bf B245}, 276
(1990).

\bibitem{L3}
L3 Collaboration: B.~Adeva {\it et al.}, Phys.\ Lett.\ {\bf B251}, 311 (1990).

\bibitem{opal}
OPAL Collaboration: M.~Z.~Akrawy {\it et al.}, Z.~Phys.\ {\bf C49}, 1 (1991).

\bibitem{rubbia}
C.~Rubbia, in {\it Proc.\ LEP-HEP\,91}, ed.\ by S.~Hegarty {\it et al.}, World
Scientific (1991), p.\,439.

\bibitem{kz}
Z.~Kunszt and F.~Zwirner, in {\it Proceedings of the Large Hadron
Collider Workshop at Aachen}, CERN 90-10; revised results in CERN-TH.6150/91.

\bibitem{680}
V.~Barger, M.~S.~Berger, A.~L.~Stange and R.~J.~N.~Phillips,
Madison report MAD/PH/680 (1991), Phys.\ Rev.\ D (in press).

\bibitem{baer}
H.~Baer, M.~Bisset, C.~Kao and X.~Tata, Florida State Univ.\
report FSU-HEP-911104 (1991).

\bibitem{gunion}
 J.~F.~Gunion, R.~Bork, H.~E.~Haber and A.~Seiden, U.~C.~Davis
report UCD-91-29; J.~F.~Gunion and L.~H.~Orr, report UCD-91-15;
J.~F.~Gunion, H.~E.~Haber, and C.~Kao, report UCD-91-32.

\bibitem{696}
V.~Barger, K.~Cheung, R.~J.~N.~Phillips, and A.~L.~Stange, {\it Collider
discovery limits for supersymmetric Higgs bosons}, UW-Madison report
MAD/PH/696.

\bibitem{janot}
P.~Janot, Orsay preprint LAL~91-61 (1991);
J.-F.~Grivaz, Orsay preprint LAL~91-63 (1991).

\bibitem{zwirner}
See for example F.~Zwirner, CERN-TH.6357/91.

\bibitem{loopcor}
S.~P.~Li and M.~Sher, Phys.\ Lett.\ {\bf B140}, 339 (1984);
J.~F.~Gunion and A.~Turski, Phys.\ Rev.\ {\bf D39}, 2701,
{\bf D40}, 2325, 2333  (1989);
M.~S.~Berger, Phys.\ Rev.\ {\bf D41}, 225 (1990);
Y.~Okada {\it et al.}, Prog.\ Theor.\ Phys.\ {\bf 85} ,1 (1991);
Phys.\ Lett.\ {\bf B262}, 54 (1991);
H.~Haber and R.~Hempfling, Phys.\ Rev.\ Lett.\ {\bf 66}, 1815 (1991);
J.~Ellis {\it et al.}, Phys.\ Lett.\ {\bf B257}, 83 (1991);
R.~Barbieri {\it et al.}, Phys.\ Lett.\ {\bf B258}, 167 (1991);
J.~Lopez and D.~V.~Nanopoulos, Phys.\ Lett.\ {\bf B266}, 397 (1991);
A.~Yamada, Phys.\ Lett.\ {\bf B263}, 233 (1991);
M.~Drees and M.~N.~Nojiri, KEK-TH-305 (1991);
R.~Hempfling, SCIPP-91/39;
M.~A.~Diaz and H.~E.~Haber, SCIPP-91/14 (1991);
D.~M.~Pierce, A.~Papadopoulos and S.~Johnson, LBL-31416 (1991);
P.~H.~Chankowski, S.~Pokorski and J.~Rosiek, Phys.\ Lett.\ {\bf B274}, 191
(1992);
J.~R.~Espinosa and M.~Quiros, Phys.\ Lett.\ {\bf B279}, 92 (1992).

\bibitem{langacker92}
P.~Langacker, in the {\it Review of Particle Properties}, Phys.\ Rev.\
{\bf D45}, VII.159 (1992).

\end{references}
\end{document}